\documentclass{article}
\usepackage{amsmath}
\usepackage{amsfonts}
\usepackage{amssymb}
\usepackage{graphicx}

\input{tcilatex}

\begin{document}

\title{From Information Geometry to Newtonian Dynamics\thanks{%
Presented at MaxEnt 2007, the 27th International Workshop on Bayesian
Inference and Maximum Entropy Methods (July 8-13, 2007, Saratoga Springs,
New York, USA).}}
\author{Ariel Caticha and Carlo Cafaro \\
%EndAName
{\small Department of Physics, University at Albany-SUNY, }\\
{\small Albany, NY 12222, USA.}}
\date{}
\maketitle

\begin{abstract}
Newtonian dynamics is derived from prior information codified into an
appropriate statistical model. The basic assumption is that there is an
irreducible uncertainty in the location of particles so that the state of a
particle is defined by a probability distribution. The corresponding
configuration space is a statistical manifold the geometry of which is
defined by the information metric. The trajectory follows from a principle
of inference, the method of Maximum Entropy. No additional \textquotedblleft
physical\textquotedblright\ postulates such as an equation of motion, or an
action principle, nor the concepts of momentum and of phase space, not even
the notion of time, need to be postulated. The resulting entropic dynamics
reproduces the Newtonian dynamics of any number of particles interacting
among themselves and with external fields. Both the mass of the particles
and their interactions are explained as a consequence of the underlying
statistical manifold.
\end{abstract}

\section{Introduction}

It is widely assumed that geometry is useful because it describes properties
of the real world. Indeed, Euclidean geometry may very well have been the
first successful physics theory, the first example of a \textquotedblleft
law of nature\textquotedblright .\ Later developments such as Riemannian
geometry and the theory of fiber bundles have only strengthened this
conception: geometry works because it lies at the very core of physics.
Thus, it may be surprising, at least at first sight, to find that the same
methods of geometry have also turned out to be useful in statistical
inference, a separate field that makes no claims to authority on natural
phenomena. It could just be a coincidence but perhaps it is not.

Perhaps the laws of physics are deeply geometrical because they are
practical rules to process information about the world and geometry is the
uniquely natural tool to do just that. This notion, that \emph{the laws of
physics are not laws of nature but rules of inference}, seems outrageous but
deserves serious attention. The evidence supporting it is already
considerable. Indeed, most of the formal structure of statistical mechanics 
\cite{Jaynes57} and of quantum theory \cite{Caticha98} can already be
derived from principles of inference (consistency, probabilities, entropy,
etc.).

The objective of this paper is to use well established principles of
inference to derive Newtonian dynamics from relevant prior information
codified into a statistical model. The challenge, of course, is to
accomplish this task without assuming what we want to derive. One must not
assume equations of motion or principles of least action, and in particular,
one must not assume the concept of momentum and the associated phase space,
and not even the notion of an absolute Newtonian time.

The first step is to construct a suitable statistical model of the space of
states of a system of particles. A most remarkable fact is that the
statistical configuration space is automatically endowed with a geometry and
that this \textquotedblleft information\textquotedblright\ geometry turns
out to be unique \cite{Amari00}\cite{Cencov81}.

Next we tackle the dynamics: Given the initial and the final states, what
trajectory is the system expected to follow? In the usual approach one
postulates an equation of motion or an action principle that presumably
reflects a \textquotedblleft law of nature.\textquotedblright\ For us the
dynamics follows from a principle of inference, the method of Maximum
Entropy, and we show that with a suitable choice of the statistical manifold
the resulting \textquotedblleft entropic dynamics\textquotedblright\ \cite%
{Caticha01}\cite{Caticha0305} reproduces Newtonian dynamics.

The entropic dynamics approach allows us to see familiar notions such as
time, mass and interactions from an unfamiliarly fresh perspective. For
example, there is no reference to an external time but there is an internal
\textquotedblleft intrinsic\textquotedblright\ time that is a measure of the
change of the system itself. Thus, the Newtonian universe turns out to be
its own clock, and the familiar Newtonian time is not particularly
fundamental but merely a convenient definition designed to make motion look
as simple as possible. Both the mass of the particles and their interactions
are explained in terms of an irreducible uncertainty of their positions;
they are features of the underlying statistical manifold.

\section{Configuration space as a statistical manifold}

Let us start with a single particle moving in space: the configuration space
is a three dimensional manifold with some unknown metric tensor $g_{ij}(x)$.
Our main assumption is that there is a certain fuzziness to space; there is
an irreducible uncertainty in the location of the particle. Thus, when we
say the particle is at the point $x$ what we mean is that its
\textquotedblleft true\textquotedblright\ position $y$ is somewhere in the
vicinity of $x$. This leads us to associate a probability distribution $%
p(y|x)$ to each point $x$ and the configuration space is thus transformed
into a statistical manifold: a point $x$ is no longer a structureless dot
but a probability distribution.

Remarkably there is a \emph{unique} measure of the extent to which the
distribution at $x$ can be distinguished from the neighboring distribution
at $x+dx$. It is the information metric of Fisher and Rao \cite{Amari00}.
Thus, physical space, when viewed as a statistical manifold, inherits a
metric structure from the distributions $p(y|x)$. We will assume that the
originally unspecified metric $g_{ij}(x)$ is precisely the information
metric induced by the distributions $p(y|x)$.

In \cite{Caticha0305} we proposed that a Gaussian model,\footnote{%
We adopt the standard summation convention: repeated indices are summed over.%
} 
\begin{equation}
p(y|x)=\frac{\gamma ^{1/2}(x)}{(2\pi )^{3/2}}\,\exp \left[ -\frac{1}{2}%
\gamma _{ij}(x)(y^{i}-x^{i})(y^{j}-x^{j})\right] ,  \label{Gaussian a}
\end{equation}%
where $\gamma =\det \gamma _{ij}$, incorporates the physically relevant
information which consists of an estimate of the particle position, 
\begin{equation}
\langle y^{i}\rangle =\tint dy\,p(y|x)\,y^{i}=x^{i}\,,  \label{exp position}
\end{equation}%
and of its uncertainty given by the covariance matrix, 
\begin{equation}
\left\langle (y^{i}-x^{i})(y^{j}-x^{j})\right\rangle =\tint
dy\,p(y|x)(y^{i}-x^{i})(y^{j}-x^{j})=\tilde{\gamma}^{ij}(x)~,
\label{exp uncertainty}
\end{equation}%
where $\tilde{\gamma}^{ij}$ is the inverse of $\gamma _{ij}$, $\tilde{\gamma}%
^{ik}\gamma _{kj}=\delta _{j}^{i}$.

Unfortunately the expected values in eqs.(\ref{exp position}) and (\ref{exp
uncertainty}) are not covariant under coordinate transformations. Indeed,
the transformation $y^{\prime i}=f^{i}(y)$ does not lead to $x^{\prime
i}=f^{i}(x)$ because in general $\langle f(y)\rangle \neq f(\langle y\rangle
)$ except when uncertainties are small. Our Gaussian model can at best be an
approximation valid when $p(y|x)$ is sharply localized in a very small
region within which curvature effects are negligible. Fortunately this is
all we need for our present purpose.

[As an interesting aside we note that it is possible to devise fully
covariant models. Here is an example: Let $\gamma _{ij}(x)$ be a positive
definite tensor field and let us use it as if it were a metric tensor, $%
d\ell ^{2}=\gamma _{ij}dx^{i}dx^{j}$. Let $\ell (x,y)$ be the $\gamma $%
-length along the $\gamma $-geodesic from the point $x$ to the point $y$.
The proposed distribution is 
\begin{equation}
p(y|x)=\frac{1}{\zeta }\gamma ^{1/2}(y)\exp -\frac{\ell ^{2}(x,y)}{2\sigma
^{2}(x)}~,  \label{covariant p}
\end{equation}%
which is a manifestly covariant object: the normalization constant $\zeta $,
the $\gamma $-length $\ell (x,y)$, the scalar field $\sigma (x)$, and $%
dy\,\gamma ^{1/2}(y)$ are all invariants. From this model we can compute a
second metric, the information metric $g_{ij}$, which need not in general
coincide with $\gamma _{ij}$. In the limit of small uncertainties (after
absorbing $\sigma $ into $\gamma _{ij}$) one recovers eq.(\ref{Gaussian a}).]

\section{The Information Metric}

The information distance between $p(y|\theta )$ and $p(y|\theta +d\theta )$
where the $\theta ^{a}$ are parameters is calculated from (see e.g., \cite%
{Amari00}) 
\begin{equation}
d\ell ^{2}=G_{ab}\,d\theta ^{a}d\theta ^{b}\quad \text{with}\quad
G_{ab}=\int dy\,p(y|\theta )\frac{\partial \log p(y|\theta )}{\partial
\theta ^{a}}\frac{\partial \log p(y|\theta )}{\partial \theta ^{b}}~.
\label{info metric}
\end{equation}%
Consider the 9-dimensional space of Gaussians 
\begin{equation}
p(y|x,\gamma )=\frac{\gamma ^{1/2}}{(2\pi )^{3/2}}\exp \left[ -\frac{1}{2}%
\gamma _{ij}(y^{i}-x^{i})(y^{j}-x^{j})\right] ~.
\end{equation}%
Here the parameters $\theta ^{a}$ include the three $x^{i}$ plus six
independent elements of the symmetric matrix $\gamma _{ij}$. Eq.(\ref{info
metric}) gives the information distance between $p(y|x,\gamma )$ and $%
p(y|x+dx,\gamma +d\gamma )$ as 
\begin{equation}
d\ell ^{2}=G_{ij}dx^{i}dx^{j}+G_{k}^{ij}d\gamma
_{ij}dx^{k}+G^{ij\,kl}d\gamma _{ij}d\gamma _{kl}~,
\end{equation}%
where 
\begin{equation}
G_{ij}=\gamma _{ij}\,,\quad G_{k}^{ij}=0\,,\quad \text{and}\quad G^{ij\,kl}=%
\frac{1}{4}(\tilde{\gamma}^{ik}\tilde{\gamma}^{jl}+\tilde{\gamma}^{il}\tilde{%
\gamma}^{jk})~.
\end{equation}%
($\tilde{\gamma}^{ik}$ is the inverse of $\gamma _{kj}$.) Therefore, 
\begin{equation}
d\ell ^{2}=\gamma _{ij}dx^{i}dx^{j}+\frac{1}{2}\tilde{\gamma}^{ik}\tilde{%
\gamma}^{jl}d\gamma _{ij}d\gamma _{kl}~.  \label{info metric 9d}
\end{equation}%
This is the metric of the full 9-dimensional manifold, but it is not what we
need.

What we want is the metric of the embedded 3-dimensional submanifold where $%
\gamma _{ij}=\gamma _{ij}(x)$ is some function of $x$. To find the induced
metric we cannot just substitute $d\gamma _{ij}=\partial _{k}\gamma
_{ij}\,dx^{k}$ into eq.(\ref{info metric 9d}) because under a change of
coordinates $dx^{k}$ transforms as a tensor but the ordinary derivative $%
\partial _{k}\gamma _{ij}$ does not. In a model of physical space the $i$
indices in $x^{i}$ cannot be treated independently from the $ij$ indices
that appear in $\gamma _{ij}$ because any transformation that changes the $%
x^{i}$ also changes the $\gamma _{ij}$. Accordingly, we require that $%
d\gamma _{ij}=\nabla _{k}\gamma _{ij}\,dx^{k}$ where $\nabla _{k}$ is the
covariant derivative and the corresponding induced information metric is 
\begin{equation}
g_{ij}=\gamma _{ij}+\frac{1}{2}\tilde{\gamma}^{ac}\tilde{\gamma}^{bd}\nabla
_{i}\gamma _{ab}\nabla _{j}\gamma _{cd}\,~.  \label{info metric a}
\end{equation}

Normally one is given a manifold of probability distributions and the
problem is to find the corresponding information metric. In order to do
physics we are also concerned with the inverse problem: we want to design
statistical manifolds with the appropriate geometries. We want to find the
covariance field tensor $\gamma _{ij}(x)$ that leads to a given metric
tensor $g_{ij}(x)$. Thus, we regard eq.(\ref{info metric a}) as a set of
differential equations for $\gamma _{ij}(x)$. Since $\nabla _{k}g_{ij}=0$,%
\footnote{%
The choice of the Levi-Civita connection is justified in the next section.}
a straightforward substitution shows that the solution is 
\begin{equation}
\gamma _{ij}(x)=g_{ij}(x)~.  \label{info metric b}
\end{equation}%
In words: \emph{information distance is measured in units of the local
uncertainty}. This beautifully simple but non-trivial result is valid in the
low uncertainty regime where eq.(\ref{Gaussian a}) holds. The uniqueness of
the solution (\ref{info metric b}), and whether it also holds in high
curvature regions, such as near singularities, remains to be ascertained.

\section{Entropic Dynamics for a single particle}

The key to the question \textquotedblleft Given an initial and a final
state, what trajectory is the system expected to follow?\textquotedblright\
lies in the implicit assumption that there exists a continuous trajectory. A
large change is the result of a succession of very many small changes and 
\emph{therefore} we only need to determine what a short segment of the
trajectory looks like. The idea behind entropic dynamics is that as the
system moves from a point $x$ to a neighboring point $x+\Delta x$ it must
pass through a halfway point \cite{Caticha01}.

The basic dynamical question can now be rephrased as follows: The system is
initially described by the probability distribution $p(y|x)$ and we are
given the information that it has moved to one of the neighboring states in
the family $p(y|x^{\prime })$ where the $x^{\prime }$ lie on the plane
halfway between the initial $x$ and the final $x+\Delta x$. Which $%
p(y|x^{\prime })$ do we select? The answer is given by the method of maximum
(relative) entropy, ME. The selected distribution is that which maximizes
the entropy of $p(y|x^{\prime })$ \emph{relative} to the prior $p(y|x)$
subject to the constraint that $x^{\prime }$ is equidistant from $x$ and $%
x+\Delta x$. The result is that the selected $x^{\prime }$ minimizes the
distance to $x$ and therefore the three points $x$, $x^{\prime }$ and $%
x+\Delta x$ lie on a straight line.

Since any three neighboring points along the trajectory must line up, the
trajectory predicted by entropic dynamics is the geodesic that minimizes the
length 
\begin{equation}
J=\tint\limits_{\lambda _{i}}^{\lambda _{f}}d\lambda \left[ g_{ij}\dot{x}^{i}%
\dot{x}^{j}\right] ^{1/2}\quad \text{with}\quad \dot{x}^{i}=\frac{dx^{i}}{%
d\lambda }~,  \label{Jacobi}
\end{equation}%
where $\lambda $ is any parameter that labels points along the curve, $%
x^{i}=x^{i}(\lambda )$.

Incidentally, note that in entropic dynamics there is one family of curves
that is singled out as special: these are the minimal-length geodesics. From
the purpose of building useful physics models no additional structure is
needed and thus none will be introduced. It is therefore natural to use this
same family of curves to \emph{define} the notion of parallelism: the
minimal-length geodesics are defined to be the straightest curves. This
definition leads to the Levi-Civita connection which is equivalent to the
condition $\nabla _{k}g_{ij}=0$ assumed in the previous section. (See e.g. 
\cite{Schutz80})

The simplest statistical model is a three-dimensional manifold of
spherically symmetric Gaussians with constant variance $\sigma _{0}^{2}$.
The corresponding information metric is 
\begin{equation}
g_{ij}^{(0)}(x)=\gamma _{ij}^{(0)}(x)=\frac{1}{\sigma _{0}^{2}}\delta _{ij}~,
\label{euclidean  metric}
\end{equation}%
which we recognize as the familiar metric of flat Euclidean space. It is
reassuring that already in such a simple model entropic dynamics reproduces
the familiar straight line trajectories that are commonly associated with
Galilean inertial motion. But this is too simple; non-trivial dynamics
requires some curvature.

We are thus led to consider a slightly more complicated model of spherically
symmetric Gaussians where the variance is a non-uniform scalar field $\sigma
^{2}(x)$. It is convenient to write the corresponding information metric as
the Euclidean metric eq.(\ref{euclidean metric}) modulated by a (positive)
conformal factor $\Phi (x)$, 
\begin{equation}
g_{ij}(x)=\gamma _{ij}(x)=\frac{\Phi (x)}{\sigma _{0}^{2}}\delta _{ij}~,
\label{1-particle metric}
\end{equation}%
with $\sigma ^{2}(x)=\sigma _{0}^{2}/\Phi (x)$.\footnote{%
The effect of $\Phi (x)$ is a local dilation. Since each side of a small
triangle at $x$ is dilated by the same factor $\Phi (x)$ its angles remain
unchanged. Such angle-preserving transformations are called conformal.}

It is convenient to rewrite the length eq.(\ref{Jacobi}) with the metric (%
\ref{1-particle metric}) in the form

\begin{equation}
J=2^{1/2}\tint\limits_{\lambda _{i}}^{\lambda _{f}}d\lambda \,L(x,\dot{x})~,
\label{Jacobi a}
\end{equation}%
with a \textquotedblleft Lagrangian\textquotedblright\ function 
\begin{equation}
L(x,\dot{x})=[\Phi (x)T_{\lambda }(\dot{x})]^{1/2}\quad \text{with}\quad
T_{\lambda }(\dot{x})=\frac{1}{2\sigma _{0}^{2}}\delta _{ij}\dot{x}^{i}\dot{x%
}^{j}~.
\end{equation}%
The geodesics follow from the Lagrange equations, 
\begin{equation}
\frac{d}{d\lambda }\frac{\partial L}{\partial \dot{x}^{i}}=\frac{\partial L}{%
\partial x^{i}}~,
\end{equation}%
or%
\begin{equation}
\frac{1}{\sigma _{0}^{2}}\left( \frac{\Phi }{T_{\lambda }}\right) ^{1/2}%
\frac{d}{d\lambda }\left[ \left( \frac{\Phi }{T_{\lambda }}\right) ^{1/2}%
\frac{dx^{i}}{d\lambda }\right] =\frac{\partial \Phi }{\partial x^{i}}~.
\end{equation}%
These rather formidable equations can be simplified considerably once we
notice that the parameter $\lambda $ is quite arbitrary. Let us replace the
original $\lambda $ with a new\ parameter $t$ given by 
\begin{equation}
dt=\left( \frac{T_{\lambda }}{\Phi }\right) ^{1/2}d\lambda \quad \text{or}%
\quad \frac{d}{dt}=\left( \frac{\Phi }{T_{\lambda }}\right) ^{1/2}\frac{d}{%
d\lambda }~.  \label{new t}
\end{equation}%
In terms of the new $t$ the equation of motion simplifies to 
\begin{equation}
\frac{1}{\sigma _{0}^{2}}\frac{d^{2}x^{i}}{dt^{2}}=\frac{\partial \Phi }{%
\partial x^{i}}~.  \label{Newton a}
\end{equation}%
From eq.(\ref{new t}) the new $t$ is such that 
\begin{equation}
\Phi =T_{\lambda }\left( \frac{d\lambda }{dt}\right) ^{2}=T_{t}\quad \text{%
where}\quad T_{t}=\frac{1}{2\sigma _{0}^{2}}\delta _{ij}\frac{dx^{i}}{dt}%
\frac{dx^{j}}{dt}~.  \label{E cons a}
\end{equation}%
Eqs.(\ref{Newton a}) and (\ref{E cons a}) are equivalent to Newtonian
dynamics. To make it explicit we introduce a \textquotedblleft
mass\textquotedblright\ $m$ and a \textquotedblleft
potential\textquotedblright\ $\phi (x)$ through a mere change of notation, 
\begin{equation}
\frac{1}{\sigma _{0}^{2}}=m\quad \text{and}\quad \Phi (x)=-\phi (x)+E\,
\end{equation}%
where the constant $E$ reflects the freedom to add a constant to the
potential. The result is Newton's equation, 
\begin{equation}
m\frac{d^{2}x^{i}}{dt^{2}}=-\frac{\partial \phi }{\partial x^{i}}~,
\label{Newton b}
\end{equation}%
and energy conservation,%
\begin{equation}
\frac{1}{2}m\delta _{ij}\frac{dx^{i}}{dt}\frac{dx^{j}}{dt}+\phi (x)=E~,
\label{E cons b}
\end{equation}%
Thus, the constant $E$ is interpreted as energy.

We have just derived $F=ma$ purely from principles of inference applied to
the relevant information codified into a statistical model! From eq.(\ref%
{Jacobi}) onwards our inference approach is formally identical to the Jacobi
action principle of classical mechanics \cite{Lanczos 70} but we did not
need to know this. Indeed, by a wild stretch of our historical imagination
it is perhaps conceivable that had Newton, Lagrange, and Jacobi known less
physics and much more inference they might have invented their subject along
these lines. Had history actually followed this unlikely course we might not
have used the notions of mass $m$ or potential $\phi (x)$ and instead we
would have referred to the particle's \textquotedblleft
intrinsic\textquotedblright\ position uncertainty $\sigma _{0}$, and how it
is modulated throughout space by the field $\Phi (x)$.

The derivation above serves to illustrate the main idea but suffers from two
important limitations. First, it applies to a single particle with a fixed
constant energy and this means that we deal with an isolated system. Second,
while it is true that we have identified a convenient and very suggestive
parameter $t$, how do we know that it actually represents \textquotedblleft
true\textquotedblright\ time? Is $t$ the universal Newtonian time or just a
parameter that applies only to one particular isolated particle? The
original formulation in terms of the \textquotedblleft
Jacobi\textquotedblright\ action, eq.(\ref{Jacobi a}), is completely
timeless; how and where did time sneak in?

The solution to both these problems emerges as we apply the formalism to the
motion of the only system known to be completely isolated: the whole
universe. Then the fact that the energy is a fixed constant does not
represent a restriction. And further, since the preferred time parameter
would be associated to the whole universe, it would not be at all
inappropriate to call it the \emph{universal} time.

\section{The whole universe: many particles}

To simplify our notation we will consider a universe that consists of $N=2$
particles. The generalization to arbitrary $N$ is trivial. For the $2$%
-particle system the position $x=(x_{1},x_{2})$ is denoted by $6$
coordinates $x^{A}$ with $A=1,2,\ldots 6$. Let $x^{A}=(x^{i_{1}},x^{i_{2}})$
with $i_{1}=1,2,3$ for particle 1 and $i_{2}=4,5,6$ for particle 2. A point
in the $N=2$ configuration space is a Gaussian distribution, 
\begin{equation}
p(y|x)=\frac{\gamma ^{1/2}(x)}{(2\pi )^{3/2}}\,\exp \left[ -\frac{1}{2}%
\gamma _{AB}(x)(y^{A}-x^{A})(y^{B}-x^{B})\right] ~.  \label{Gaussian b}
\end{equation}%
The simplest model for two (possibly non-identical) particles assigns
uniform variances $\sigma _{1}^{2}$ and $\sigma _{2}^{2}$ to each particle.
The corresponding metric, analogous to eq.(\ref{euclidean metric}), is 
\begin{equation}
g_{AB}^{(0)}=\gamma _{AB}^{(0)}=m_{AB}~,  \label{euclidean metric b}
\end{equation}%
where $m_{AB}$ is a constant $6\times 6$ diagonal matrix, 
\begin{equation}
m_{AB}=%
\begin{bmatrix}
\delta _{i_{1}j_{1}}/\sigma _{1}^{2} & 0 \\ 
0 & \delta _{i_{2}j_{2}}/\sigma _{2}^{2}%
\end{bmatrix}%
~,
\end{equation}%
where each entry represents a $3\times 3$ matrix. The metric $m_{AB}$
describes a flat space; the trajectories are familiar \textquotedblleft
straight\textquotedblright\ lines and the particles move independently of
each other; they do not interact. As before, non-trivial dynamics requires
the introduction of curvature and the simplest way to do this is through an
overall conformal field $\Phi (x)$ with $x=(x_{1},x_{2})$. Thus we propose 
\begin{equation}
g_{AB}(x)=\gamma _{AB}(x)=\Phi (x)m_{AB}~.  \label{2-particle metric}
\end{equation}%
The equation of motion for the $N=2$ universe is the geodesic that minimizes 
\begin{equation}
J=2^{1/2}\tint\limits_{\lambda _{i}}^{\lambda _{f}}d\lambda \,L(x_{1},x_{2},%
\dot{x}_{1},\dot{x}_{2})~,  \label{Jacobi b}
\end{equation}%
where 
\begin{equation}
L(x,\dot{x})=[\Phi (x)T_{\lambda }(\dot{x})]^{1/2}\quad \text{and}\quad
T_{\lambda }(\dot{x})=\frac{1}{2}m_{AB}\dot{x}^{A}\dot{x}^{B}~.
\end{equation}%
The Lagrange equations yield, 
\begin{equation}
m_{AB}\left( \frac{\Phi }{T_{\lambda }}\right) ^{1/2}\frac{d}{d\lambda }%
\left[ \left( \frac{\Phi }{T_{\lambda }}\right) ^{1/2}\frac{dx^{B}}{d\lambda 
}\right] =\frac{\partial \Phi }{\partial x^{A}}~,
\end{equation}%
which suggests introducing a new parameter $t$ defined by 
\begin{equation}
dt=\left( \frac{T_{\lambda }}{\Phi }\right) ^{1/2}d\lambda \quad \text{or}%
\quad \frac{d}{dt}=\left( \frac{\Phi }{T_{\lambda }}\right) ^{1/2}\frac{d}{%
d\lambda }~.  \label{time t}
\end{equation}%
In terms of the new parameter the equations of motion are 
\begin{equation}
m_{AB}\frac{d^{2}x^{A}}{dt^{2}}=\frac{\partial \Phi }{\partial x^{A}}~,
\end{equation}%
which, since $m_{AB}$ is a diagonal matrix, is 
\begin{equation}
\frac{1}{\sigma _{n}^{2}}\frac{d^{2}x^{i_{n}}}{dt^{2}}=\frac{\partial }{%
\partial x^{i_{n}}}\Phi (x_{1},x_{2})~,
\end{equation}%
for each of the particles, $n=1,\,2$. Note that the motion of particle 1
depends on the location of particle 2: \emph{these are interacting particles!%
}

The new time parameter $t$, eq.(\ref{time t}), is such that 
\begin{equation}
\Phi =T_{\lambda }\left( \frac{d\lambda }{dt}\right) ^{2}=T_{t}\quad \text{%
where}\quad T_{t}=\frac{1}{2}m_{AB}\frac{dx^{A}}{dt}\frac{dx^{B}}{dt}~.
\end{equation}%
As before, the equivalence to Newtonian dynamics is made explicit by a
change of notation, 
\begin{equation}
\frac{1}{\sigma _{n}^{2}}=m_{n}\quad \text{and}\quad \Phi (x)=-\phi (x)+E\,~.
\end{equation}%
The result is 
\begin{equation}
m_{n}\frac{d^{2}x^{i_{n}}}{dt^{2}}=-\frac{\partial }{\partial x^{i_{n}}}\phi
(x_{1},x_{2})\quad \text{and}\quad \frac{1}{2}m_{AB}\frac{dx^{A}}{dt}\frac{%
dx^{B}}{dt}+\phi (x_{1},x_{2})=E~.
\end{equation}%
The constant $E$ is the total energy of the universe and there are no
restrictions on the energy of individual subsystems.

For the conformal factor $\Phi (x_{1},x_{2})$ we can choose anything we
want. For example, 
\begin{equation}
\Phi (x_{1},x_{2})=-v_{1}(x_{1})-v_{2}(x_{2})-u(x_{1},x_{2})+E~,
\end{equation}%
so the particles can interact with external potentials $v_{1}$ and $v_{2}$
and also with each other through $u(x_{1},x_{2})$.

The definition of time $t$ required taking into account all the particles in
the universe. This is in accord with the ephemeris time defined by
astronomers. We started with a completely timeless theory, eq.(\ref{Jacobi b}%
), and in fact, no \emph{external }time has been introduced. What we have is
a convenient $t$ parameter associated to the change of the total system,
which in this case is the whole universe. The universe is its own clock; it
measures universal time. Incidentally, note that the reparametrization that
allowed us to introduce a Newtonian time was possible only because the same
conformal factor $\Phi (x)$ applies equally to all particles.

Entropic dynamics offers a new perspective on the concepts of mass and
interactions. To see this note that since $\gamma _{AB}$ is diagonal the
distribution (\ref{Gaussian b}) turns out to be a product, 
\begin{equation}
p(y|x)=p(y_{1}|x_{1},x_{2})p(y_{2}|x_{1},x_{2})~.
\end{equation}%
Note that although the model represents interacting particles the
distribution is a product: the uncertain variables $y_{1}$ and $y_{2}$ are
statistically independent. The coupling arises through conditioning on $%
x=(x_{1},x_{2})$.

Let us focus our attention on particle 1; similar remarks also apply to
particle 2. The distribution $p(y_{1}|x_{1},x_{2})$ is a spherically
symmetric Gaussian, 
\begin{equation}
p(y_{1}|x_{1},x_{2})\propto \,\exp \left[ -\frac{1}{2\sigma
_{1}^{2}(x_{1},x_{2})}\delta _{ij}(y^{i}-x^{i})(y^{j}-x^{j})\right] ~.
\end{equation}%
The uncertainty in the position of particle 1 is given by 
\begin{equation}
\sigma _{1}(x_{1},x_{2})=\left[ \Phi (x_{1},x_{2})m_{1}\right] ^{-1/2}.
\end{equation}%
The mass $m_{1}$ is interpreted in terms of a uniform background
contribution to the uncertainty. Mass is a manifestation of an uncertainty
in location; higher mass reflects a lower uncertainty. On the other hand,
interactions arise from the non-uniformity of $\sigma _{1}(x_{1},x_{2})$
that depends on the location of other particles through the modulating field 
$\Phi (x_{1},x_{2})$. It is worthwhile to note that even though this is a
non-relativistic model there already appears a \textquotedblleft
unification\textquotedblright\ between mass and (potential) energy: they are
different aspects of the same thing, the position uncertainty.

\section{Final remarks}

We emphasize that the model we have proposed does not take into account all
the dynamical information that we know is relevant -- relativistic and
quantum effects have not been included. Our model is very restricted. For
example, our model invokes two apparently unrelated metrics. There is the
metric $\delta _{ij}$ of flat 3-dimensional Euclidean space that appears in
the kinetic energies and there is the information metric $g_{ij}$ that
accounts for mass and interactions and applies to the curved configuration
space. This is a reflection of the fact that a system of $N$ particles is
described as a point in a $3N$-dimensional configuration space. A better
model would have $N$ points living within the same evolving 3-dimensional
space.

Furthermore, we have not provided any rationale for how to choose the
modulating field $\Phi (x)$. Just as Newton deliberately refrained from
explaining the origin of his inverse square forces -- \emph{hypothesis non
fingo }-- so have we refrained from offering any physical hypothesis about
the underlying fuzziness of space. It is reasonable to expect that a
derivation of general relativity as an example of entropic dynamics would
yield important insights on this matter. Preliminary steps in this direction
appeared in \cite{Caticha0305}.

What we have done is to show, by exhibiting an explicit example, that the
tools of inference -- probability, information geometry and entropy -- are
sufficiently rich that one can construct entropic dynamics models that
reproduce recognizable laws of physics. Perhaps all laws of physics can be
derived in this way.


\begin{thebibliography}{9}
\bibitem{Jaynes57} E. T. Jaynes: Phys. Rev. \textbf{106}, 620 and \textbf{108%
}, 171 (1957); \emph{E. T. Jaynes: Papers on Probability, Statistics and
Statistical Physics}, ed. by R. D. Rosenkrantz (Reidel, Dordrecht, 1983).

\bibitem{Caticha98} A. Caticha: Phys. Lett. \textbf{A244}, 13 (1998); Phys.
Rev. \textbf{A57}, 1572 (1998); Found. Phys. \textbf{30}, 227 (2000)
(arXiv.org/abs/quant-ph/9810074); \textquotedblleft From Objective
Amplitudes to Bayesian Probabilities\textquotedblright\ in \emph{Foundations
of Probability and Physics-4},\emph{\ }ed. by G. Adenier, C. Fuchs, and A.
Khrennikov, AIP Conf. Proc. Vol. 889, 62 (2007)
(arXiv.org/abs/quant-ph/0610076).

\bibitem{Amari00} S. Amari and H. Nagaoka, \emph{Methods of Information
Geometry} (Am. Math. Soc./Oxford U. Press, Providence, 2000).

\bibitem{Cencov81} N. N. \v{C}encov: \emph{Statistical Decision Rules and
Optimal Inference}, Transl. Math. Monographs, vol. 53, Am. Math. Soc.
(Providence, 1981); L. L. Campbell: Proc. Am. Math. Soc. \textbf{98}, 135
(1986).

\bibitem{Caticha01} A. Caticha, \textquotedblleft Entropic
Dynamics\textquotedblright\ in \emph{Bayesian Inference and Maximum Entropy
Methods in Science and Engineering}, ed. by R. L. Fry, AIP Conf. Proc. 
\textbf{617}, 302 (2002). (arXiv.org/abs/gr-qc/0109068).

\bibitem{Caticha0305} A. Caticha, \textquotedblleft Towards a Statistical
Geometrodynamics\textquotedblright\ in \emph{Decoherence and Entropy in
Complex Systems} ed. by H.-T. Elze (Springer Verlag, 2004)
(arXiv.org/abs/gr-qc/0301061); \textquotedblleft The Information geometry of
Space and Time\textquotedblright\ in \emph{Bayesian Inference and Maximum
Entropy Methods in Science and Engineering}, ed. by K. Knuth, A. Abbas, R.
Morris, and J. Castle, AIP Conf. Proc. \textbf{803}, 355 (2006)
(arXiv.org/abs/cond-mat/0508108).

\bibitem{Schutz80} B. F. Schutz, \emph{Geometrical Methods of Mathematical
Physics }(Cambridge U. Press, 1980).

\bibitem{Lanczos 70} C. Lanczos, \emph{The Variational Principles of
Mechanics} (Dover, New York, 1986).
\end{thebibliography}
\end{document}